\documentclass{jkas}

\def\year{2020} 
\def\month{April} 
\def\volume{00} 
\def\issue{0} 
\def\beginpage{1} 
\def\endpage{99} 
\def\received{December 13, 2019} 
\def\accepted{February 10, 2020} 
\date{Received \received; accepted \accepted}

\def\eg{{\it e.g.,} }

\def\ha{H{\rm$\alpha$}}
\def\hb{H{$\beta$}}
\def\hi{H\,{\sc i}}
\def\hii{H\,{\sc ii}}
\def\he1{He\,{\sc i}}
\def\heii{He\,{\sc ii}}

\def\caii{Ca\,{\sc ii}}

\def\niii{N\,{\sc iii}}

\def\kms{km\hspace{1pt}s$^{-1}$}
\def\cc{cm$^{-3}$}

\title{
The Formation of the Double Gaussian Line Profiles of the Symbiotic Star AG Peg
}

\author[]{Siek~Hyung}
\author[]{Seong-Jae~Lee }

\affil{Department of Earth Science Education (Astronomy), Chungbuk National University, Chungbuk, 28644, Korea; \email{hyung@chungbuk.ac.kr, seong@chungbuk.ac.kr}}

\def\corrauthor{
S.-J. Lee
}

\def\runningauthor{
Hyung \& Lee
}

\def\runningtitle{
Line Profile Formation of AG Peg
}

\def\keywords{
stars: variable: symbiotic:  individual: AG Peg --  ISM: lines and kinematics
}

\def\abstracttext{
We analyze high dispersion emission lines of the symbiotic nova AG Peg, observed in 1998, 2001, and 2002. The {\ha} and {\hb} lines show three components, two narrow and one underlying broad line components, but most other lines, such as {\hi}, {\he1}, and {\heii} lines, show two blue- and red-shifted components only.
A recent study by Lee \& Hyung (2018) suggested that the double Gaussian lines emitted from a bipolar conic shell is likely to form Raman scattering lines observed in 1998. In this study, we show that the bipolar cone with an opening angle of 74$^{o}$, which expands at a velocity of 70 {\kms} along the polar axis of the white dwarf, can accommodate the observed double line profiles in 1998, 2001, and 2002. We conclude that the emission zone of the bipolar conic shell, which formed along the bipolar axis of the white dwarf due to the collimation by the accretion disk, is responsible for the double Gaussian profiles.
}

\begin{document}
\jkashead 


\section{Introduction\label{sec:intro}}

A symbiotic star is known to be a binary system due to the observed incoherent spectral characteristics with relatively low and very high temperature. However, it is not easy to confirm the presence of two stars in imaging studies. AG Peg was first reported as a Be star by \citet{fl07}, after which \citet{me16, me42} observed He and {\caii} absorption lines, numerous ionized or neutral emission lines, and TiO$_2$ absorption bands. Considered a Cyg-Be type star at the time of discovery, AG Peg is now known as a symbiotic system composed of the Wolf-Rayet type of white dwarf (WD) and the M3 type of giant star (GS) \citep{me16, hu75}.

The spectroscopy of AG Peg in ongoing research showed a variation of its brightness and spectral intensities over several months or decades. The orbital period is about 800 days. The temperature of the WD is about 100\,000 K, while its cold component is a GS \citep{al80, vi88, nu92}.  Since the beginning of astronomical observations, the second major outburst took place in 2015, which was shorter than the first major explosion observed in 1850.

In a symbiotic binary system, the variation of the emission line profiles might be involved with the shielding effect partly due to the relative position of the observer in the Earth against the rotating binary system. Some earlier studies noted the presence of double Gaussian profiles or an absorption line at the center that might be associated with the GS passage at a specific phase \citep{co97}. However, the double Gaussian profiles seemed to exist in all phases and the GS cannot always be in front of the emission zone surrounding the WD. Lee \& Hyung (2018, hereafter LH18) showed that such a GS blocking could not form observed Gaussian line profiles. For the spectroscopic data of $\phi \sim$ 0.0 or 0.5, seen during Lick observatory observations, LH18 investigated whether the GS  blocks the emission zone around the hot WD, causing observed Gaussian line profiles. They provided some kinematic inference that the GS passages were not the source of the double Gaussian line profiles. Instead, LH18 demonstrated that a bipolar conic shell formed double Gaussian lines and that a broad line wing was formed by the Raman scattering process in the neutral region of the shell.

Raman scattering processes in symbiotic stars were proposed first by \citet{sc89}, and Raman wings were first discussed by \citet{nu89} and further studied by \citet{le02}. \citet{lh00} also showed that it could form in a dense planetary nebula such as IC 4997.

Symbiotic stars have been considered a candidate for a type Ia supernova. The kinematic structure, including the physical conditions of the ionized hydrogen region and the symbiotic binary system, is of great interest. The mass inflow from the GS into the hot component might form an accretion disk. The fast stellar wind from the hot star is likely to develop bipolar conic shells that are responsible for both the double and broad emission lines.

As the orbital phase changes, since the relative positions of the binary components also change relative to the observer, the shape of line profiles changes accordingly. LH18 suggested that the bipolar conic shell is the most likely source of double Gaussian line profiles, but they did not compute theoretical line profiles at another period. In this study, we try to fit the {\ha} or {\hb} double Gaussian line profiles observed in arbitrary phases based on the bipolar conic model geometry.

Section~\ref{sec:lick} describes the observed spectra for three epochs acquired by the Hamilton Echelle Spectrograph at Lick Observatory. Section~\ref{sec:double} presents the predicted line profiles and the best parameters of a bipolar conic shell suitable for three epochs. We investigate the geometric structure of the system, responsible for the double Gaussian line profiles, observed in 1998, 2001, and 2002 epochs. The theoretical line profiles, accountable for the different phases, are provided to accommodate the observed double line profiles. We conclude in Section~\ref{sec:con}.

\section{Lick Observatory Spectral Data\label{sec:lick}}

Spectroscopic data were obtained by Lawrence H. Aller and Siek Hyung in 1998, 2001, and 2002 at the Lick observatory in the USA using the high-resolution Hamilton Echelle Spectrograph (HES) attached to a 3 m Shane telescope. The throughput of the HES is R $\sim$ 50\,000. The slit width of the HES 640 micron used corresponds to 1.2$''$ on the sky, and the wavelength resolution is 0.1 -- 0.2\AA/pixel. The observed spectral range is 3470 -- 9775\AA, and we selected only the H and He line among the various emission lines. Long exposure times of 1800 sec and 3600 sec were chosen for weak lines in observing runs. Meanwhile, short exposure times of 300 sec and 180 sec  were also obtained to avoid saturation of the strong {\ha} line. In observations, one of the IRAF standard stars, HR 7596, was secured for later flux calibration.

Table~\ref{tab:tab1} shows the observation dates of AG Peg, Greenwich universal time (UT), Julian day, and phase. The phase of AG Peg is defined as zero when WD is to the observer in front of the GS. The 2002 data (phase $\phi$ = 11.98, 0.98 hereafter), close to $\phi$ = 0.0, were secured when the WD is in front of the GS. The 2001 data (phase $\phi$ = 11.56, 0.56 hereafter) is close to $\phi$ = 0.5 when the GS is in front of the WD, a possible GS position of blocking the emission zone near the WD. In this study, we also used the 1998 data (phase $\phi$ = 10.24, 0.24 hereafter) near 0.25 when both the WD and GS were located side by side. Detailed explanations for the 1998 spectral data were given in LH18.

\begin{table}[t!]
\caption{AG Peg observation log\label{tab:tab1}}
\centering
\begin{tabular}{ccc}
\toprule
Observation date (UT) &	Julian date (d)	& phase ($\phi$)  \\
\midrule
1998/09/17 & 	2451073.70	& 10.24  \\
2001/08/30 &	2452151.60	& 11.56  \\
2002/08/11 &	2452498.92	& 11.98  \\
\bottomrule
\end{tabular}
\end{table}

The data were analyzed using the IRAF (Image Reduction and Analysis Facilities) of the National Optical Astronomy Observatory (NOAO), and the final flux data with wavelength (\AA), corrected for atmospheric extinction, were obtained. The wavelength was converted to speed ({\kms}) so that the kinematic characteristics were immediately visible. We further converted the observed wavelength frame to the system frame, correcting the radial velocity of AG Peg relative to the Sun, -7.09 {\kms}, obtained from a large number of permitted {\hi} lines, namely the Balmer and the Paschen lines.

The final spectrum was further decomposed to find the kinematic subcomponents responsible for the full line profile, which involves a trial and error process that requires choosing the number of line components at the observer's discrimination. We used StarLink/Dipso created by Interactive Data Language (IDL) and European Southern Observatory (ESO) for this deconvolution.

\begin{figure*}[t]
\centering
\includegraphics[width=170mm]{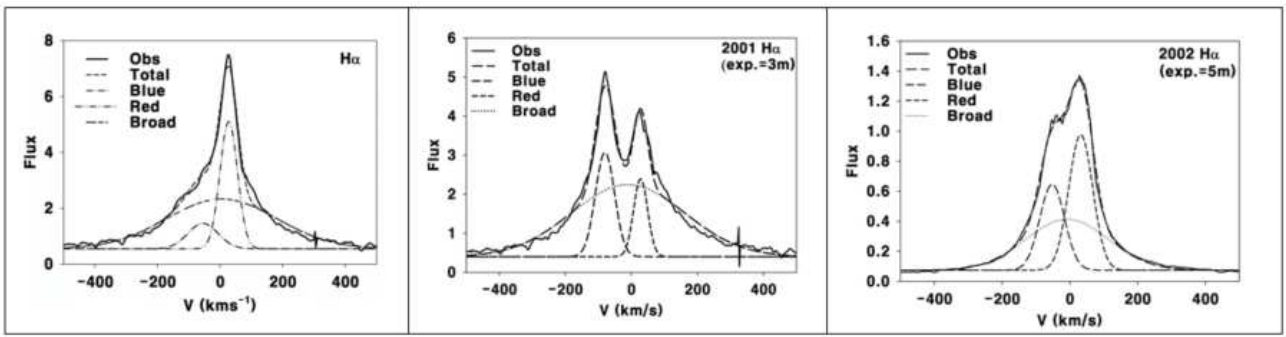}
\caption{{\ha} spectral line profiles for the 1998 ($\phi$=0.24), 2001 ($\phi$=0.56), and 2002 ($\phi$=0.98) observations from left to right, respectively. Exposures are 300 sec, 180 sec, and 300 sec, respectively. Radial velocity unit (x-axis): {\kms}. Flux unit (y-axis): 10$^{-11}$erg s$^{-1}$cm$^{-2}$ per (\kms). The kinematic center of 0 {\kms} is adapted from the LH18 result.\label{fig:jkasfig1} }

\end{figure*}

Figure~\ref{fig:jkasfig1} shows the {\ha} lines observed in three epochs. To avoid saturation in the spectral line profiles, we used the 3 min and 5 min exposures. LH18 interpreted that the {\ha} and {\hb} hydrogen lines observed in 1998 are composed of three parts, that is, double components + broad wing component, while the previous studies assumed them to be four to five pieces of different kinematical regions of, \eg, 60, 120, 400, and 1000 {\kms} \citep{ke93, er04}.

If one looks at the observed line profiles with bare eyes, the line profiles appear to consist of only two components with uncertain outlines. However, the analysis of IDL separates them into three components. The detailed explanations for the three elements were given in LH18 based on the 1998 data. The top narrow lines consist of blue- and red-shifted double Gaussian components. The {\ha} line observed in 1998 shows that the red component is stronger than the blue one, while the same line in 2001 shows that the blue component is stronger than that of the red one.

All the observed line profiles in Figure~\ref{fig:jkasfig1} show the two top narrow red-shift and blue-shift components and one  broad  component underneath. The FWHM (Full Width at Half Maximum) of the top narrow lines shown in Figure~\ref{fig:jkasfig1} is 60 -- 120 {\kms} for each element, while the width of the broad line is 400 –- 500 {\kms}. We adopted the kinematical center from the result of LH18, which was found in the strong lines observed in 1998. One expects to see {\heii} 6560 as well. Due to the relatively strong {\ha} strength, however, the {\heii} 6560 line component(s) to the left of the {\ha} line cannot be separated.

\begin{figure*}[t]
\centering
\includegraphics[width=170mm]{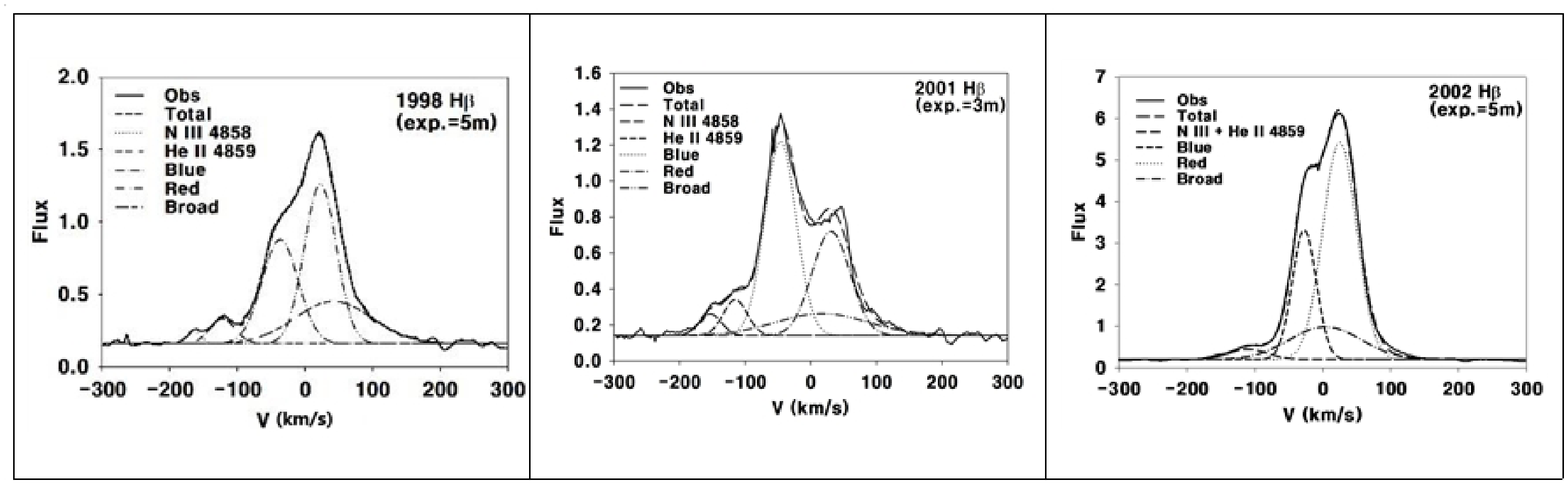}
\caption{{\hb} spectral line profiles for the 1998 ($\phi$=0.24), 2001 ($\phi$=0.56), and 2002 ($\phi$=0.98) observations, from left to right, respectively. See LH18 for the 1998 {\hb} line profile. The center of the broad line observed in 1998 is about 45.0 {\kms}, red-shifted from those in 2001 and 2002. See Figure~\ref{fig:jkasfig1} for others not mentioned.\label{fig:jkasfig2}}
\end{figure*}

The {\hb} spectral line profiles in Figure~\ref{fig:jkasfig2} show two other components to the left side beside the {\hb} components, which turned out to be {\niii} 4858 and {\heii} 4559 lines. Although the FWHMs of the {\hb} line profiles are slightly different from those of the {\ha} spectral line profiles, both consist of three components.

The interstellar extinction correction has been applied to the observed {\ha} and {\hb} fluxes with the interstellar extinction coefficient C [=log I({\hb})/F({\hb})]=0.04 [or color excess E(B−V)=0.027] adopted from Kim \& Hyung (2008, hereafter KH08). The interstellar extinction corrected I({\ha})/I({\hb}) ratio for the (blue$+$red) flux is 2.87 for 1998, and those for 2001 and 2002 have similar values, 2.97 and 2.86, respectively. The spectral lines observed in 2001 show that the red-shift intensities of the {\ha} and {\hb} fluxes are weak compared to the blue ones: the GS with the expanded atmosphere may have blocked part of the emission zone.

Inconsistencies in flux intensity or line width exist in the {\ha} and {\hb} fluxes observed in 1998 and 2002. LH18 already noted their disagreement. Meanwhile, the {\ha} and {\hb} fluxes observed in 2001 do not show such a severe discordance. The discordance might be partly involved with complex kinematics near the inner Lagrangian point and the GS with the expanded atmosphere. When the GS blocks the WD at $\phi$ $\sim$ 0.5, it also hides the inner Lagrangian point L1 and the expanded atmosphere of the GS close to the WD. As a result, both the {\ha} and {\hb} lines are not likely to be affected much by the multiple zones.

It is not straightforward to determine the kinematics center because three components are involved. However, based on the broad component of Raman scattering, the center of the broad line observed in 1998 seems to be about 45.0 {\kms} red-shift, compared with the 2001 and 2002 centers (see Figure~\ref{fig:jkasfig2}). If the emission zones are gravitationally bounded to the WD, the kinematic center would be red-shifted at $\phi$ $\sim$ 0.25 relative to others at $\phi$ $\sim$ 0.0 or 0.5 (See Figure~\ref{fig:jkasfig4}).

Table~\ref{tab:tab2} lists the measurements of the observed {\ha} and {\hb} lines for three observing runs. Note also that the ratio of the weak blue-ward flux to the strong red-ward flux is 33 : 67 (at $\phi$ $\sim$ 0.24) and 35 : 65 (at $\phi$ $\sim$ 0.98), respectively, whereas their strengths are reversed when the GS is in front, where the ratio of the strong blue-ward and the weak red-ward flux is 59 : 41 at $\phi$ $\sim$ 0.5. Such a reversal intensity variation is likely to be caused by the relative positions of the two components to the observer.

\begin{table*}[t]
\caption{Observed {\ha} and {\hb} fluxes for 3 observations\label{tab:tab2}}
\centering
\begin{tabular}{ccccccc}
\toprule
Date ($\phi$) &	Line &	Blue & Red & Raman wing	& blue : red (\%)	& mean  (\%)   \\
\midrule
1998/09/17 & {\ha}	& 1.09(--11) & 3.20(--11)	& 8.15(-−11) & 25 : 75 &    \\ 	
(10.24)   & {\hb} & 5.95(-12) & 9.07(-12) & 3.82(-12) & 40 : 60 & 33 : 67 \\
2001/08/30 & {\ha}	& 2.29(-11)	& 1.33(-11) & 6.16(-11) & 59 : 41 &   \\
(11.56)   & {\hb} & 7.18(-12) & 5.00(-12) & 7.53(-13) & 59 : 41 & 59 : 41  \\	
2002/08/11 & {\ha}	& 5.17(-11)	& 7.99(-11)	& 1.08(-10)	& 39 : 61 & \\	
(11.98)	 & {\hb}	& 1.43(-11)	& 3.18(-11)	& 5.57(-12)	& 31 : 69& 35 : 65 \\
\bottomrule
\end{tabular}
\tabnote{
Flux unit: 1.03($−$11) denotes 1.03$\times$10$^{-11}$ erg$^{-1}$s$^{-1}$cm$^{-2}$. See LH18 for the errors in these measurements.
}
\end{table*}

The {\ha} and {\hb} line profiles also consist of three  components. LH18 confirmed that both {\ha} and {\hb} lines observed in 1998 formed in the ionized gas ({\hii}) region of the H atom number density of $\sim10^{9.8}$ {\cc}. LH18 showed that the broad line with the FWHM of 400 -- 500 {\kms} is the result of the Raman scattering mechanism.

As seen in other Balmer profiles by \citet{le17}, the Raman scattering mechanism does not play a role in the {\heii} or other {\hi} lines, probably because the Lyman lines related to these lines are relatively weak in the gas densities mentioned above. The present study considers the ionized bipolar cone extended from the WD (as in LH18) and confirms that such a cone geometry accommodates the observed double Gaussian line profiles. To do this, first, we need to know whether the GS (or expanded atmosphere) blocks the receding cone when the former is in front of the WD. We will provide supporting arguments that can be hopefully determined by examining the double Gaussian line profiles.

\section{Double Gaussian Line Profiles from the Bipolar Cone\label{sec:double}}

\begin{figure*}[t]
\centering
\includegraphics[width=120mm]{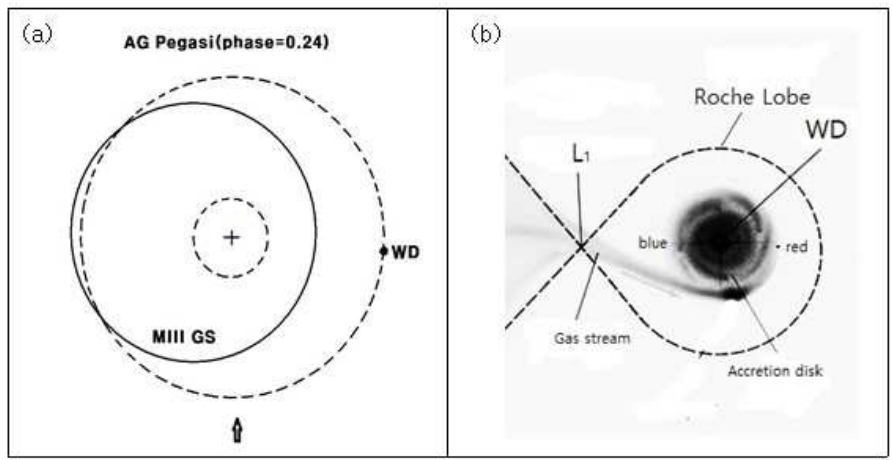}
\caption{A close binary system (a face-on view). The observer is at the bottom (arrow) when $\phi$ $\sim$ 0.25. (a) The expanded atmosphere of the GS could be the high number density of the neutral zone responsible for the Raman scattering zone. Dashed circles: orbital paths. +: the center of mass (CM). (b) The transferring gas from the GS into the WD would form an ionized accretion disk around the hot WD. Dashed line: Roche lobe. L1: the inner Lagrangian point.\label{fig:jkasfig3}}
\end{figure*}

The close binary structure is a probable geometry responsible for the Raman scattering mechanism due to the close distance of the GS atmosphere to the WD. Figure~\ref{fig:jkasfig3} schematically shows the size and relative position of both companions, assuming that the AG Peg is a close binary system. A circular orbit with an eccentricity of e = 0 is considered, and the observer is at the bottom at $\phi$ = 0.24 in the 1998 observation. The photons emitted in the ionized region would enter into a neutral hydrogen region of the GS with the expanded atmosphere where the Raman scattering occurs \citep{he15}.

\begin{figure*}[t]
\centering
\includegraphics[width=145mm]{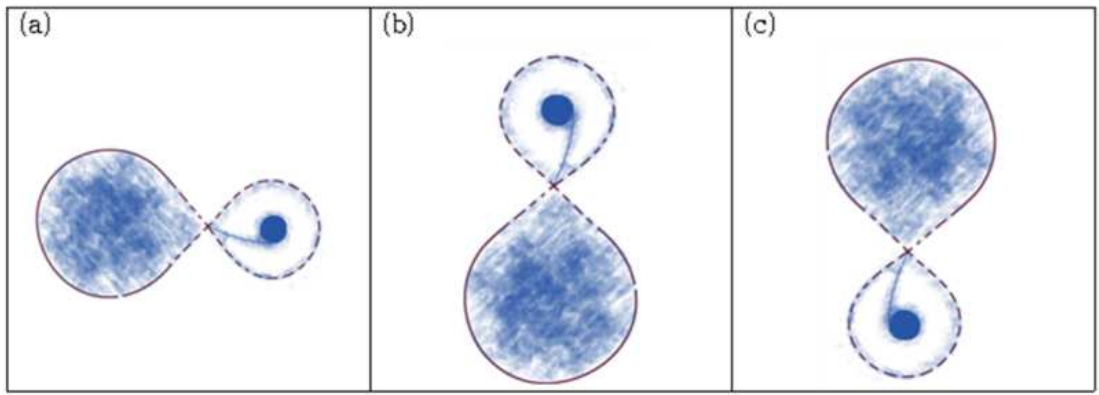}
\caption{The relative position of the WD (+accretion disk) and the GS in a semi-detached binary system (face-on view). (a) $\phi$ $\sim$ 0.24: Both stars can be seen at the same time. (b) $\phi$ $\sim$ 0.5: The WD and the accretion disk are partly blocked depending on the inclination of the line of sight. (c) $\phi$ $\sim$ 0.0: The WD and the accretion disk are not blocked to an observer. The observer is at the bottom of the figure, whose line of sight is tilted a little relative to the figure plane due to the inclination. See Figure~\ref{fig:jkasfig5}.\label{fig:jkasfig4}}
\end{figure*}

As seen in Figure~\ref{fig:jkasfig3}, the observer at the bottom can observe the spectra simultaneously coming from two stars, i.e., approaching the GS and receding the WD at $\phi$ $\sim$ 0.25. The spectral data for $\phi$ = 0.56 observed in 2001 correspond to the case where the observer is on the left side in Figure~\ref{fig:jkasfig3} (in case of counter clock-wise revolution).
The sweeping movement of the GS and the WD at $\phi$ $\sim$ 0.0 or 0.5 is tangential to the observer's radial direction on Earth.
One will not be able to observe the emission line profiles at all if the large GS blocks the WD and ionizing gas. However, the observer can see some part of the bipolar cone emitting area when the GS with a certain tilt can not block the ionizing gas completely. We have come to the conclusion that a semi-detached binary system is a better characterization for this system than a closed binary system to account for the situations. Figure~\ref{fig:jkasfig4} shows the relative positions of the GS and the WD in a semi-detached binary system with the observer at the bottom.

The overflowing gas flows to the WD through L1, filling some of the WD-side Roche lobe (See Figure~\ref{fig:jkasfig3}(b)). The size of Roche lobe in the GS is  about 20\% larger than the radius of the GS itself.
The strong stellar wind with a terminal velocity of $\sim$1000 {\kms} is believed to be produced from the hot WD due to the slow nova eruption beginning in 1850 \citep{er04}. Hence, one may assume the presence of a mass-loaded strong stellar wind from the WD, which would push the outer shell gas outward and form the bipolar cones.

\begin{figure}[t]
\centering
\includegraphics[width=75mm]{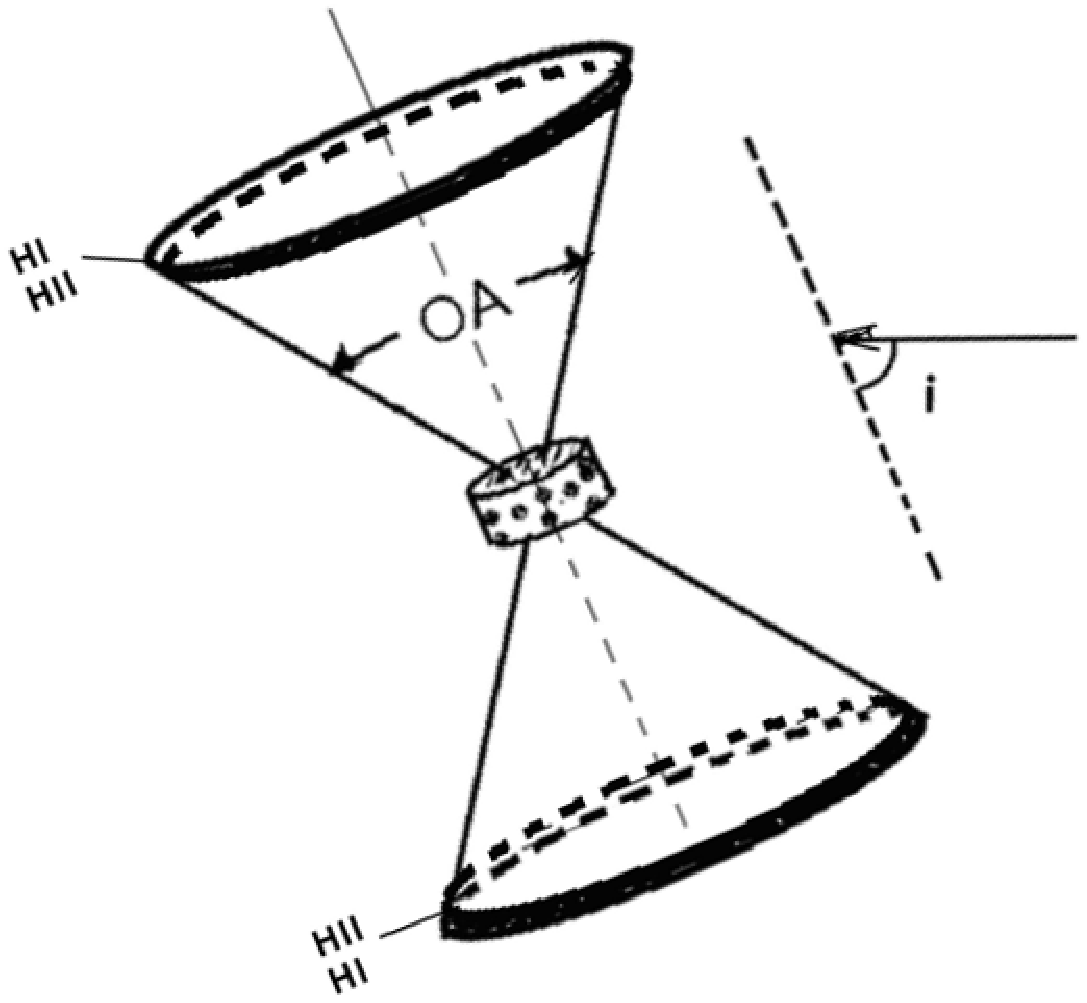}
\caption{Bipolar conic shells and accretion disk around the WD. The bipolar cones are formed due to the mass-loaded outflow from the WD against the common envelope (CE). OA: the opening angle of the cone. i: the inclination angle of the pole relative to the observer on the right side. `HII' indicates the radiation bounded emission zone in the conic shell and  `HI' indicates the neutral zone. See the text.\label{fig:jkasfig5}}
\end{figure}

Figure~\ref{fig:jkasfig5} shows a schematic diagram of the bipolar conic outflow, originating from the compact accretion disk around the WD. `HII' indicates the ionized ({\hii}) zone by the hot WD star, while `HI' indicates the neutral zone that surrounds the {\hii} zone. \citet{ke93} and \citet{er04} presumed the presence of the fast colliding emission region (between the WD and the GS) where the fast wind with a velocity of $\sim$700 {\kms} from the WD collides with the wind with a velocity of $\sim$60 {\kms}  from the GS, forming the composite line profiles with $\sim$200 {\kms} doublets. However, LH18 showed that such a broad line is actually a Raman scattering line, due to the nearby high-column density of the {\hi} zone. The `HI' and `HII' indicate that both {\hi} and {\hii} shells may exist within the conic shell, whose boundary is defined by the outer radius of the inner ionized {\hii} zone due to the shortage of UV photons from the WD.

The photoionization (P-I) model by KH08 indicates that the temperature of the WD is about 100\,000 K. Such a high WD temperature would produce fast stellar winds to be able to load the mass. The expanding outflows from the hot WD would interact with some parts of the CE and push them outwards, forming the expanding ionized bipolar cones with velocities ranging from 50 -- 100 {\kms} in the opposite.

\begin{figure*}[t]
\centering
\includegraphics[width=120mm]{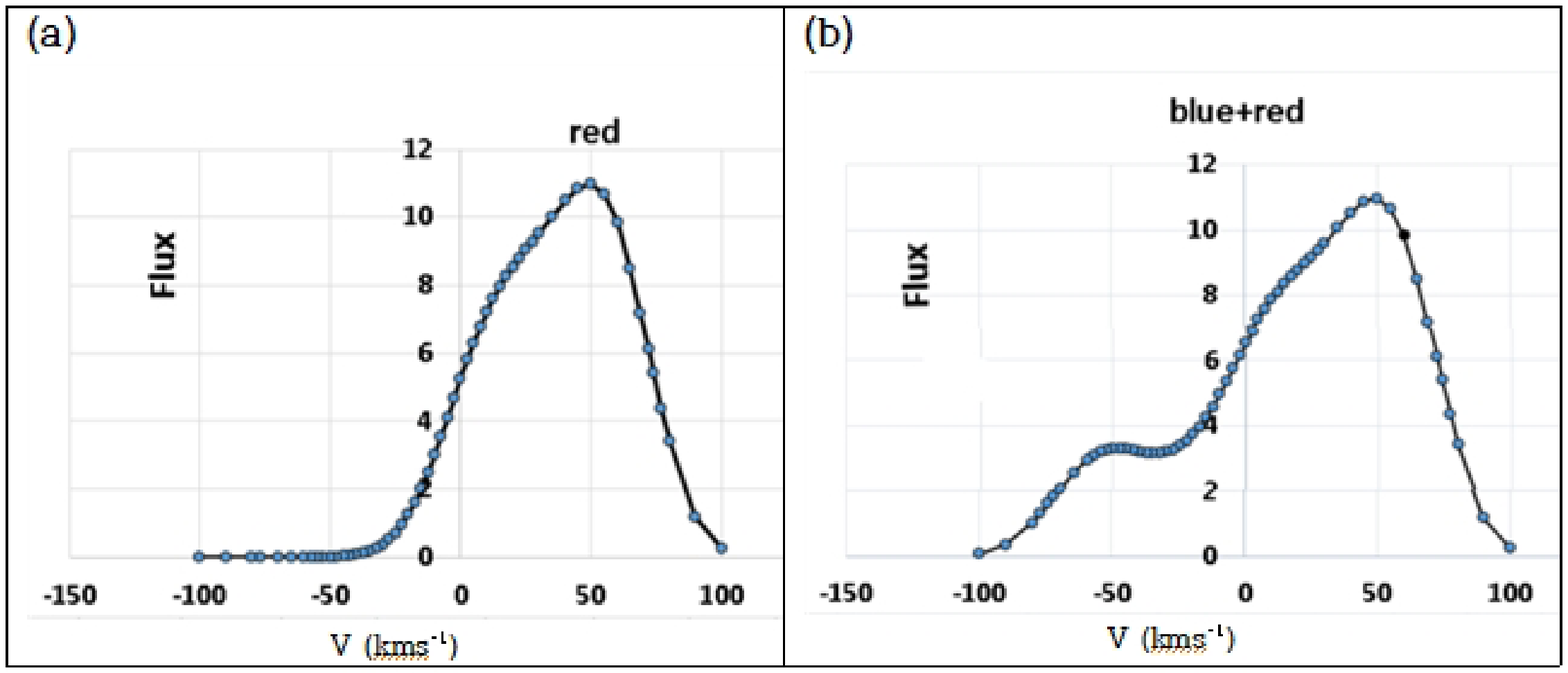}
\caption{Synthetic line profiles for the line profile observed in 1998 ($\phi$ $\sim$ 0.25). Radial velocity unit (x-axis): {\kms}. Flux unit (y-axis): (unscaled) erg s$^{-1}$ cm$^{-2}$ per ({\kms}). (a) The red component only (the same as Figure 8 in LH18). (b) A synthetic double Gaussian line profile with weaker (23\%) blue + stronger (77\%) red components. AG Peg is assumed to be a semi-detached binary system.  Note that the center of broad line in 1998 is red-shifted by about 45.0 {\kms} relative to the 2001 and 2002 centers at about 0.0 {\kms} (see Figure~\ref{fig:jkasfig2}). Hence, the center of the line at $\phi$ $\sim$ 0.25 would correspond to $\lambda$ = 6563.80{\AA} for the {\ha} line or 4862.06{\AA} for the {\hb} line.\label{fig:jkasfig6}}
\end{figure*}

Figure~\ref{fig:jkasfig6} shows the result of the theoretical double Gaussian line profiles calculated for a bipolar conic shell with the physical parameters in Table~\ref{tab:tab3}. The physical parameters are from Case A of Table 7 by LH18. The bipolar conic shells have the {\hii} emission zone responsible for the optical emission lines that are surrounded by the outer neutral {\hi} zone. The bipolar conic shell is almost hollow with a thin {\hii} zone whose inner and outer radii are 3.16 $\times$ 10$^{13}$ cm and 3.18 $\times$ 10$^{13}$ cm ($\sim$2.1AU), respectively (see Model I by LH18).

Since the bipolar conic shells expand radially outwards from the WD, one can observe the double peak line profiles. The top conic shell would mostly recede (i.e., the red-shift component of the double peak line profiles) relative to the observer at the right side in Figure~\ref{fig:jkasfig5}. In contrast, the bottom conic shell would appear to be approaching toward us (i.e., the blue-shift component). The opening angle of the cone, OA, would determine the line width. In contrast, the inclination of the polar axis i, would determine the degree of the peak separation of the double Gaussian profiles.

The expansion speed of the bipolar cones pushed by the stellar wind of the hot WD was assumed to be 70 {\kms}. The opening angle is OA = 74$^{\circ}$, and the inclination of the polar axis relative to the observer on Earth is i = 55$^{\circ}$, which would give the FWHM of 53 {\kms}, similar to the observed 44 –- 55 {\kms}. Although the diagram assumes that both conic shells are the same, their size or opening angle could be different, e.g., the small bottom of the conic shell against the large top conic shell. In fitting the synthetic profile, we applied a Gaussian smoothing factor $\sigma_{\rm G}$ = 15 {\kms} to the artificial profile, smaller than the expected intrinsic line broadening factor, $\sim$45 {\kms}.

The conic shell is related to the Roche lobe of the WD or CE, which expands along the polar axis of the rotating WD. The radial distance of the photoionized zone is larger than the radius of the GS or the orbital path, so the emission zone responsible for the observed emission lines would not be blocked by them.

\begin{table}[t!]
\caption{Model parameters for 1998\label{tab:tab3}}
\centering
\begin{tabular}{cc}
\toprule
Parameters &	 1998  \\
\midrule
V$_{\rm exp}$ ({\kms}) &	70   \\
Opening angle, OA ($^{\circ}$) &	74   \\
Inclination angle, i ($^{\circ}$)	& 55   \\
FWHM ({\kms})	& 53              \\
Observed FWHM ({\kms}) &	44 -- 55     \\
$\sigma_{\rm G}$({\kms})   & 	15               \\
\bottomrule
\end{tabular}
\end{table}

The synthetic double Gaussian profiles in Figure~\ref{fig:jkasfig6}(b) consist of the weaker (23\%) blue- and the stronger (77\%) red-shifted components, due to the small shell volume fraction of the lower conic shell and the more substantial fraction of the upper conic shell. The opening angle of the cone, OA = 74$^{\circ}$, could be also smaller in the lower conic shell.

The bipolar cones expanding in opposite directions would form two peaks, whose separation varies depending on the inclination angle of the pole. From various trials, we found that the inclination axis of the pole, i = 55$^{\circ}$, gives double Gaussian peaks separated by $\pm$37 {\kms} suitable for the observation. At this stage, we do not yet consider a possible occultation by the GS. The relatively large GS could eclipse some portion of the CE and the conic shell at a specific line of sight.

As seen in Figures~\ref{fig:jkasfig1} and~\ref{fig:jkasfig2}, the blue-shift components of the {\hi} line profiles observed in 1998 were weaker than the red-shift ones. The blue-shifted versus red-shifted flux intensity ratios from the mean of the {\ha} and {\hb} fluxes were 33\% : 67\% (1998) ($\phi$ $\sim$ 0.25), which changed as 59\% : 41\% (2001) (at $\phi$ $\sim$ 0.5). The red component became weaker in 2001 (near $\phi$ $\sim$ 0.5), suggesting that the expanded atmosphere of the GS or GS with CE blocks the receding portion of the conic shell and the WD itself. However, the red-shifted Gaussian component again became stronger in 2002 ($\phi$ $\sim$ 0.0).

Some earlier studies, e.g., \citet{nu95}, \citet{ke93}, and \citet{er04}, assumed various emission zones located between the GS and the WD to accommodate such a semi-periodic variation. However, such an analysis may easily fall into an ad hoc modification, without a self-consistent geometrical structure provided. The weaker flux intensities in the blue-shifted line profiles and the stronger flux intensities in the red-shifted line profiles appear to be intrinsic, originating from the geometry itself. Hence, the reversal flux ratio at $\phi$ $\sim$ 0.5 might be due to the eclipsing effect of the emission zone by the expanded atmosphere of the GS.

\begin{figure*}[t]
\centering
\includegraphics[width=160mm]{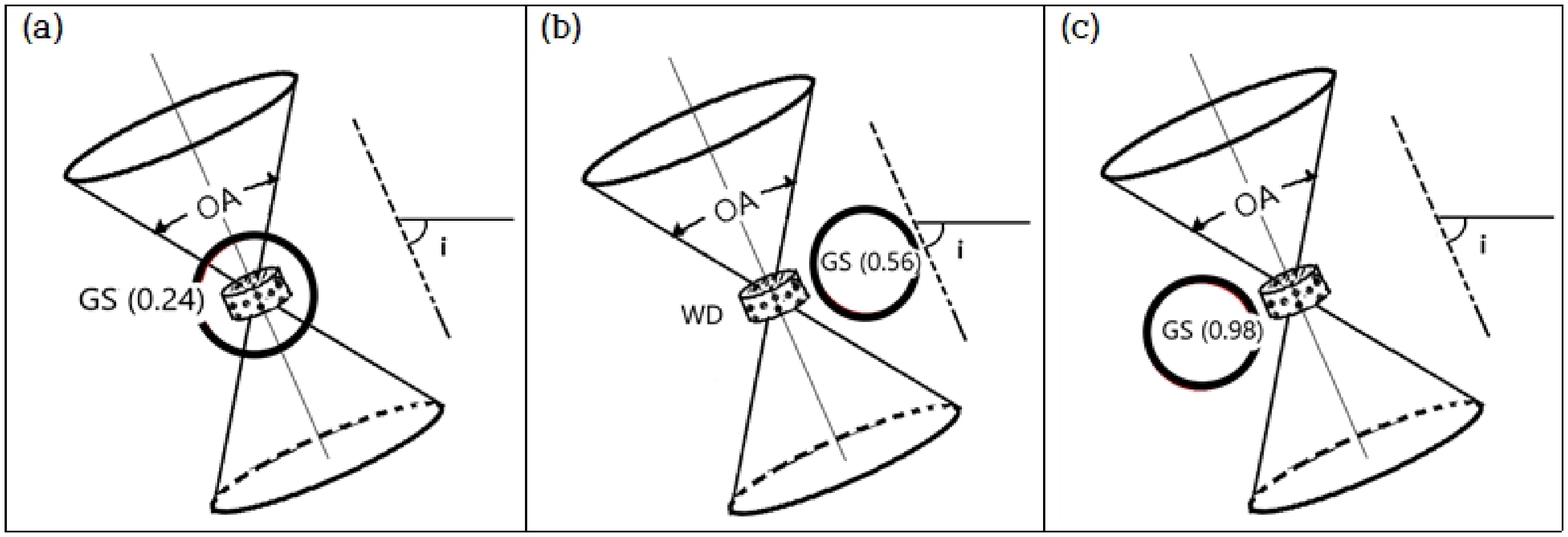}
\caption{
Bipolar conic shells, the accretion disk around the WD, and the GS (the large circle in the center). The observer (or the line of sight) is on the right side. (a) The apparent position of the GS and the accretion disk at $\phi$ $\sim$ 0.24. (b) The GS and the accretion disk (WD) relative to the line of sight at $\phi$ $\sim$ 0.56. (c) The GS and the accretion disk (WD) at $\phi$ $\sim$ 0.98. Note that the observer at $\phi$ $\sim$ 0.24 and 0.98 can see the conic shells and the accretion disks. In contrast, the expanded atmosphere of the GS at $\phi$ $\sim$ 0.56 would hide some part of the receding top conic shell. The proposed bipolar cones may be a part of the WD with the CE. See also Figure~\ref{fig:jkasfig4} for the relative positions of two stars and the CM.\label{fig:jkasfig7}
}
\end{figure*}

Figure~\ref{fig:jkasfig7} shows the bipolar conic shells along with the relative position of the GS (circle) and an accretion disk around the WD. It does not clearly show the center of mass (CM) of the revolution of the AG Peg binary system. The observer is on the right side of the figures. See Figures~\ref{fig:jkasfig3} and~\ref{fig:jkasfig4} for the center of mass (CM) and the relative positions of both stars. The {\hi} and {\hii} zones are not specified in this figure. Still, we assume that the {\hii} zone is relatively small enough, so its inner radius is comparable to the size of the expanded atmosphere of the GS.

In Figure~\ref{fig:jkasfig7}(a),  the GS is slightly above the accretion disk plane (or above the figure plane) at the phase of $\phi$ $\sim$ 0.24, while Figure~\ref{fig:jkasfig7}(b) and Figure~\ref{fig:jkasfig7}(c) show their relative positions at the phases of $\phi$ $\sim$ 0.56 and 0.98, respectively.  Note that the GS at $\phi$ $\sim$ 0.24 and 0.98 would not block the conic shell for an observer on the right side. Meanwhile, the GS (or its expanded atmosphere) would hide some part of the receding top conic shell in a case when the GS with the expanded atmosphere or the GS with the CE is large enough at $\phi$ $\sim$ 0.56.

Figure~\ref{fig:jkasfig7} indicates that the {\hii} zone is gravitationally bound to the revolving WD. As mentioned in Figure~\ref{fig:jkasfig2}, the kinematic center of the line profiles observed in 1998 is red-shifted by about +45.0 {\kms} compared to the other two line profiles.
In the predicted line profiles, we have not refined such a orbital motion.
Although the apparent position of the GS at $\phi$ = 0.24 overlaps with the WD and the accretion disk in the figure, the observer on the right side can see the GS, the {\hii} zone, and the accretion disk.

\begin{figure*}[t]
\centering
\includegraphics[width=130mm]{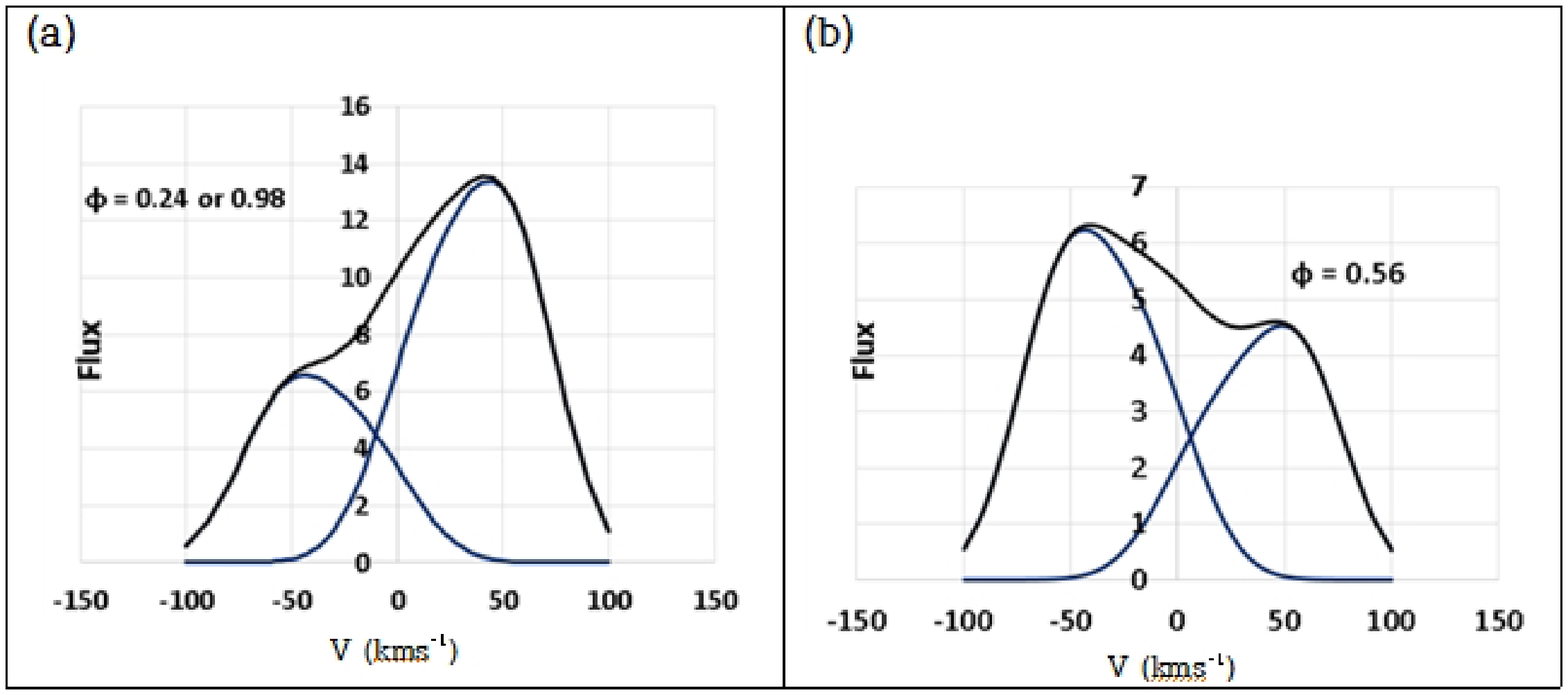}
\caption{Predicted line profiles. The flux intensities are unscaled. (a) The predicted line profiles for $\phi$ $\sim$ 0.24 or $\phi$ $\sim$ 0.98. (b) The predicted line profiles for $\phi$ $\sim$ 0.56.   0 {\kms} corresponds to 6562.82{\AA} for the {\ha} line or 4861.33{\AA} for the {\hb} line. We have not considered revolutionary motion. See Figures~\ref{fig:jkasfig2} and~\ref{fig:jkasfig6}.\label{fig:jkasfig8}}
\end{figure*}

Figure~\ref{fig:jkasfig8} shows the predicted line profiles for three phases. Figure~\ref{fig:jkasfig8}(a) represents the line profile for both $\phi$ $\sim$ 0.25 (1998) and $\phi$ $\sim$ 0.0 (2002). The physical scale of the bottom conic shell could be smaller than that of the top one (not depicted in the model geometry of Figure~\ref{fig:jkasfig7}). The contribution of the bottom and top cone shells differs, to fit the observed weaker (35\%) blue-shifted and stronger (65\%) red-shifted line profile components, respectively.

Note the position of the GS and the {\hii} conic shell at $\phi$ $\sim$ 0.98 and $\phi$ $\sim$ 0.24 in Figure~\ref{fig:jkasfig7}, where the GS does not block the {\hii} zone completely. Meanwhile, Figure~\ref{fig:jkasfig8}(b) shows the synthetic line profile with a stronger (60\%) blue component and a weaker (40\%) red component (for 2001). Hence, we conclude that the bipolar cone mainly consists of a strong upper conic structure and a weaker lower substructure. The radius of the GS appears to be smaller than the inner radius of the {\hii} zone shell (R$_{\rm i}$
$\sim$ 2 AU). However, the radius of the expanded atmosphere (R$_{\rm e}$) or the Roche lobe of the GS in a position between the {\hii} zone and the observer could hide a significant fraction of the {\hii} zone in the top conic shell due to the relatively large inclination angle: i = 55$^{\circ}$, i.e., R$_{\rm i}$  $<$  R$_{\rm e}$ $\times$ (1 + tan (90 - i$^{\circ}$)).

As seen in Figures~\ref{fig:jkasfig1} and~\ref{fig:jkasfig2}, the observed double Gaussian peaks exhibit separation at -37 and +37 {\kms}, and accordingly, the predicted peaks have the same separation. We applied a more realistic Gaussian smoothing factor of $\sigma_{\rm G}$ = 20 {\kms} for the synthetic line profiles in Figure~\ref{fig:jkasfig8}.

Table~\ref{tab:tab4} summarizes the parameters for the bipolar conic shell geometry in Figure~\ref{fig:jkasfig7}, responsible for the theoretical line profiles in Figure~\ref{fig:jkasfig8}. Note that the parameters in Table~\ref{tab:tab4} are the same as those in Table~\ref{tab:tab3}. The only difference is the relative position of the GS and the WD (not specified) and the Gaussian smoothing factor.

\begin{table}[t!]
\caption{Final model parameters\label{tab:tab4}}
\centering
\begin{tabular}{cc}
\toprule
Parameters &	 1998, 2001, 2002 \\
\midrule
V$_{\rm exp}$ ({\kms}) &	70   \\
Opening angle, OA ($^{\circ}$) &	74   \\
Inclination angle, i ($^{\circ}$)	& 55   \\
FWHM ({\kms})	& 53              \\
Observed FWHM ({\kms}) &	44 -- 60 (55)     \\
$\sigma_{\rm G}$({\kms})   & 	20             \\
\bottomrule
\end{tabular}
\end{table}

\begin{figure*}[t]
\centering
\includegraphics[width=175mm]{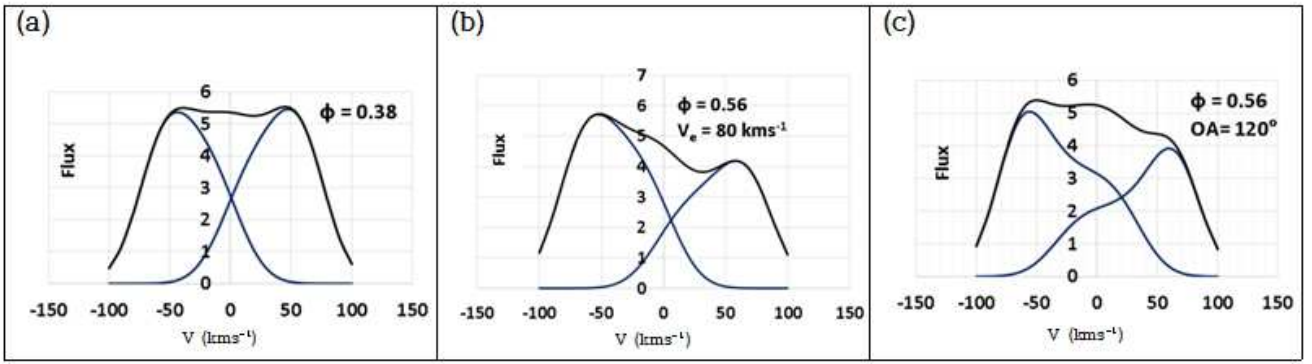}
\caption{Predicted line profiles with different kinematic parameters. Radial velocity unit (x-axis): {\kms}. Flux unit (y-axis): (unscaled) erg s$^{-1}$ cm$^{-2}$ per ({\kms}). (a) $\phi$ $\sim$ 0.35 -- 0.40 represents an arbitrary point when the GS partially blocks the {\hii} zone.  (b) $\phi$ $\sim$ 0.56 for the case of faster expansion. (c) $\phi$ $\sim$ 0.56, when the opening angle of the bipolar cone is larger, i.e., OA = 120$^{\circ}$. See Figure~\ref{fig:jkasfig6} for others not mentioned.\label{fig:jkasfig9}}
\end{figure*}

Figure~\ref{fig:jkasfig9} shows the theoretical profiles: (a) for an arbitrary binary orbital phase where both double components show the same intensity, (b) for the different expansion velocity of the shell, and (c) for a wider opening angle. The chosen inclination appears to be appropriate for the observed peak separation, so we did not vary the inclination angle. The first line profile in Figure~\ref{fig:jkasfig9}(a) corresponds to the profiles at $\phi$ $\sim$ 0.35 -- 0.40, while the other two profiles in Figures~\ref{fig:jkasfig9}(b) and (c) are calculated at $\phi$ $\sim$ 0.56. Note that the large opening angle not only widens the line profiles but also produces the apparent three components, suitable for more than two multiple line profiles secured in other observations.

The emission line formed on the hot accretion disk may be an X-ray with a blue + red component (see Figure~\ref{fig:jkasfig7}), which is not discussed in our study. Earlier studies, e.g., \citet{nu95}, \citet{ke86}, \citet{ke93}, and \citet{er04}, also identified other emission regions. (1) X-ray emission harder than a few 100 eV was detected, which might be given off by a hot plasma with a temperature of a few million K shock-heated by colliding winds. (2) The blue-shifted absorption zone from the wind regions and the surrounding nebula also exists, affecting the UV lines and producing P Cygni profiles. The former could be related to the accretion disk or the shock interaction between two stars, while the latter could be partly due to the hot WD with the stellar wind and the cool GS with the expanded atmosphere (see Figures~\ref{fig:jkasfig4} and~\ref{fig:jkasfig7}).

\section{Conclusions\label{sec:con}}

Our analysis showed that the expanding bipolar conic shell could accommodate the observed {\ha} and {\hb} double Gaussian lines, assuming a semi-detached binary system as the most probable structure in AG Peg. The observed {\hi} spectral line profiles are reasonably well described by the expanding bipolar conic shells having a polar axis with an inclination angle of i = 55$^{\circ}$ and an opening angle of OA = 74$^{\circ}$. The expansion velocity of the shell is likely to be about 70 {\kms}.

The emission zone is probably the result of the interaction of the fast stellar wind from WD and the inflow gas from the GS. The fast stellar wind must be confined to a bipolar cone direction by the dense accretion disk structure around the WD. The thick accretion disk itself must be the highly excited zone responsible for the X-ray emission observed in other studies. As shown by LH18, the Lyman lines can also be formed in similar double Gaussian profiles, which will transform into the broad Balmer {\ha} and {\hb} lines through the Raman scattering process in the outer {\hi} thick shell.

Our proposed bipolar conic shells could accommodate the observed double Gaussian profiles (FWHM $\sim$ 60 {\kms}), which are also responsible for Raman scattered broad lines ($\sim$ 400 -- 500 {\kms}). The size of the upper conic shell appears to be different in size from the lower one. However, when the GS is in front along the line of sight, the upper conic shell is liable to be blocked by the passage of the GS. The next fate of the expanding gas from a bipolar shape can be inferred from a large sale radio image, which still shows a bipolar shape.

The {\hii} zone is a radiation bounded thin shell, surrounded by the relatively thick high-density of {\hi} shell. KH08 showed that the gas number density of the {\hii} zone is n$_{\rm H}$ = 10$^9$ -- 10$^{10}$ {\cc}, while LH18 showed that the outer {\hi} zone has a column density of n$_{\rm H}$ = 3 -- 5 $\times$ 10$^{19}$ cm$^{-2}$. Beyond the outer boundary of {\hi} zone, there would be a relatively low-density of {\hi} in bipolar zones (of 20$''$ -- 1$'$ in scale), which has been confirmed by the 1.5 GHz and 5 GHz radio studies \citep{ke91}.

We conclude that the  emission zone of the bipolar conic shell, responsible for double Gaussian profiles, was formed along the polar axis of the WD due to the collimation by the accretion disk around the WD.
As more observational data accumulates through continuous monitoring in future observations, detailed analysis can lead to more concrete conclusions.


\acknowledgments

This research was supported by a grant from the National Research Foundation of Korea (NRF2015 R1D1A3A01019370; NRF2017 R1D1A3B03029309). We owe a debt of gratitude to the late Professor, Lawrence H. Aller of UCLA who joined the HES observation at the Lick Observatory for this study. We express our gratitude to Dr. Lee, K.-H. who helped Figures~\ref{fig:jkasfig1} and~\ref{fig:jkasfig2}. We also thank the anonymous referees for reviewing this paper and Eugenia H. for proof-reading.



\end{document}